\documentclass[11pt]{article}
\begin{document}
\textheight 8.3in
\textwidth 9in
\oddsidemargin 0.5in
\topmargin -0.5in
\parindent 0in
\begin{center}

{\bf ``But since the affairs of men rest still uncertain, let's reason with the worst that may befall''}\footnote{
Wm. Shakespeare, {\it Julius Caesar}, Act 5, Sc. 1}

Probability, risk, and the 2009 L'Aquila Earthquake\\
\medskip

Steven N. Shore\\
Dipartimento di Fisica ``Enrico Fermi'', Universit\'a di Pisa\footnote{To be published in 2012, {\it ScienzaePace}, 
the journal of the Interdisciplinary Center for Peace Studies (Centro Interdisciplinare Scienze per la Pace), University of Pisa  \\  ISSN 2039-1749 (http://scienzaepace.unipi.it/ ).  Sincere thanks to Giorgio Gallo, Fabio Del Sordo, Ivan De Gennaro Aquino, Alessandro Cilla, and Matteo Cantiello for valuable comments and discussions.}\\

13 Nov. 2012
\end{center}
\bigskip

\parindent 0.2in

\section{Introduction}

This article is a commentary on the verdict of the ``L'Aquila Six'', the group of bureaucrats and scientists  tried by an Italian court as a result of their public statements in advance of the quake of 2009 Apr. 6 that left the city in ruins and cause more than 300 deaths.\footnote{Detailed analyses of the seismic event is provided by D'Agostino, N. et al. 2012, {\it J. Geophys. Res.}, {\bf 117}, B02402;  Gallovic, F. and Zaradnik, J. 2012, {\it J. Geophys. Res.}, {\bf 117}, B04307; Ameri, G. et al. 2012, {\it J. Geophysical Res.}, {\bf 117}, B04308.  These together provide a comprehensive summary of the event from the geophysical side.  For general information about earthquakes and analysis, see the US Geological Survey Earthquakes Hazard Program website.} It was not the worst such catastrophic event in recent Italian history, but it {\it was} one of -- if not the -- worst failures of risk assessment and preventive action.  The six were found guilty and condemned by a first level of the justice system to substantial prison terms.  The outcry provoked by the verdict in the world press and the international scientific community has fueled the already fiery debate over whether the six should have been tried at all.  They have been presented  as martyrs to science being treated as scapegoats by a scientifically illiterate justice system and inflamed local population for not being able to perform the impossible (predict the event).  Petitions of support have been drafted and signed by thousands of working scientists and technical experts in many fields excoriating the court and the country for such an outrage against the scientific community, often accompanied by ominous warnings about the chilling effect this will have on the availability of expert advice in times of need.

My purpose here, to state it at the outset, is to explain why this view of the events of the trial is misguided, however well intentioned, and misinformed.  

To begin with the same affirmation that began the trial, that of the public prosecutor ({\it procura}), everyone knows you can't predict an earthquake, especially one like this.  That was not the motive for the indictment.  The six -- and the head of civil defense at the time -- were guilty of making false, deliberately misleading statements that were intended to calm the public in a time of increasing alarm and that caused imprudent actions -- incautious actions -- based on their {\it expert statements}.   The case has one benefit, however: it has shown a spotlight on a fundamental issue that the academic and research communities have often ignored -- the responsibility to communicate clearly, frequently, {\it and directly} with the public at large.

\section{Probability and risk in brief}

Reporters, both immediately before and even long (years) afterward, bandied about quantifications related to the l'Aquila earthquake and its likelihood.   In many reports quoted a probability for this event, most cited a value of less than 0.001 percent.  But let's examine, for a moment, what this means  -- both for the prediction itself and for the impression such a quantification gave to an uninformed public.  

\subsection{Probabilities of events}
   
If you deal a deck of cards, or throw dice, the events are independent (assuming honesty on the part of the player) and therefore these probabilities can be predicted by simple frequencies, the number of combinations that {\it can} occur and the number of ways a particular hand, or throw, {\it might} devolve.  For most people, this is as close as they ever get to probabilistic reasoning.\footnote{A cure for this is Jaynes, E. T. 2003, {\it Probability Theory: The Logic of Science} (Cambridge: Cambridge Univ. Press).}

  The same is true for lotteries, or any other form of {\it honest} gambling; it suffices to know the frequencies.  But that assumes two things.  First, that the distribution is complete (the information is complete, for example, when you know how many cards you have in a deck) {\it and} that the underlying process is a set of independent, completely random events that will occur with the same distribution each time.  

On the other hand, consider the physical world -- not the  casino.  An earthquake is a complex series of events, not simply a piece of rock that waits until the ``proper moment'' to slip out of place.   Each event is conditioned by the history of each locale, and being an intrinsically complex structure, the crust of this planet is quite variable in its mechanical and chemical properties.  The same rock, subject to different conditions of tension, groundwater, heating, and large scale stresses (e.g. plate tectonics, the movements of the large scale crust)  will have different reactions to small changes.  More important, each piece is in contact with its surroundings whose properties may not be known with such pinpoint accuracy as, say, the location of a card in a shuffle (for instance, done by a good magician).  The Earth has been surveyed by geologists in more or less detail, depending on the location, but only for the past century.  Except when a deeper portion has been exposed, often by a violent event such as a landslide, earthquake, or mountain building, or probed by wells or mining, the subsurface is only a reconstruction based on the ``boundary''.  In some cases, this can be done with enough certainty that predictions can be made of the presence of faults and cracks, rupture zones, and stressed regions within hundreds of meters.  In others, the number of samples covering a surface sampling is much coarser and only reveals structures tens of kilometers in size.   The deeper you want to go, the more ``fuzzy'' the picture becomes.   So you don't necessarily even know what cards are in the shuffle when you try to predict a hand.  

Now we come to the next, more serious problem.   The so-called ``gambler's fallacy'' is the belief that a string of recurrences is informative -- the same card appearing in successive deals or a coin that has been tossed with a run of identical results -- increases the chance that the {\it next} event will be the same.  But in a {\it random} process that is {\it independent} that is not true.  You still have the same {\it frequency-dependent} outcome, it's simply that the run has not yet been long enough to see the alternate outcomes.  That is {\it not} the case with many triggered events.   Just because you have a probability of, say, 50\% that a coin will come up heads does not mean you can't have a run of twenty heads.  It is simply unlikely for a fair coin.  But the best bet for the next toss should still be based on the 50\% assumption.  This is what is called, in statistics, a {\it Gaussian process}.  It has a mean value and a range, and that's it.   On that basis you can predict what the chance of a particular string of events might be but not any single event: {\it there is no history or memory}.  Rocks are different!  They have a ``memory'', the deformations and alterations of the past change the behavior for the future and this leads to extreme events happening with far higher chance than would be expected from a Gaussian process.  The example proffered by economists for this is the high frequency of stock market crashes (the bread-and-butter of {\it econophysics})  but there is a lesson to be transferred to earthquakes and other natural catastrophes.  If something has occurred in the past, it changes the likelihood that it may occur in the future.  This is {\it not} a probability that is well understood, either at the technical level -- it is a central question for a  very large range of disciplines -- not by the public at large.  

\subsection{The perception and assignment of risk}

Ultimately, any calculation or analysis that produces a quantitative interval of confidence in a prediction, passes to a second level -- the interpretation or qualitative assessment of risk -- when viewed in the advisory context.    Note that here we enter the murky territory of {\it perception}.\footnote{Some excellent commentaries on this issue, directed to engineers by engineers, are the books by Samuel Florman and Henry Petrosky.}

What may seem highly unlikely to one person may, instead, seem inevitable to another.  The {\it translation} is no different than between languages,.  To be effective, it must capture the sense, the essential content, without distortion but it never is -- nor can it ever be -- unique.  In general, the on of the weakest aspects of Western education is the almost complete lack of statistical reasoning.  Not that the public isn't awash in ``statistics'' , every human activity from calcio to politics to economics is ``quantified'' by the media and, by exposure and acquired habit, to individuals.  But the {\it meaning} of both these quantifications and their attendant uncertainties is rarely, if ever, explained.   We will return to this point in the next section.

In assessing risk for natural catastrophic events, the details are devilishly important and these {\it cannot} be completely known.  That is why models are used for predictions, and why the assumptions of the models have to embody the relevant uncertainties.  One never, for instance, runs one atmospheric model to predict weather, it is an ensemble of models and a range of inputs (i.e. measurements, real data) with all of their associated measurement uncertainties.  To be precise, when a meteorologist cites a ``chance of rain'' it means that a range of model runs gave that outcome and/or that the approximate conditions will occur that have, in the past, produced such an event.  Such assessments may be purely statistical and historical (although for meteorology they are not now on a daily basis but are on the seasonal or annual level)  but for earthquakes they must be much more often based on the historical data.\footnote{For a good example, related to hurricane risk assessment and communication, see Demuth, J. L. et al. 2012, {\it Bull. Amer. Meteor. Soc.}, {\bf 93}, 113.  The same applies to tornado warnings and other extreme weather, and the validity of long-term weather forecasting.}

It is this that makes the public statements  particularly delicate.  They cannot be absolute, they are based on evaluation of contextualized data.  A single event may be meaningless, a train of events will have far more information, and a comparison with historical series becomes even more informative.  The hypotheses of risk are, consequently, modified as more information comes in and cannot be made definitive.  Instead, and this is another lesson learned from the weather, {\it trends} are especially significant.  In the L'Aquila quake, the normal seismicity of an active region was known from monitoring for decades.  A few extreme events in that case, not clustered in time, would be a cause for increased attention but might -- with an emphasis on {\it might} -- not indicate anything special.  But, if there is a trend, or a cluster, the possibility that the conditions could be changing cannot be excluded and the caution level should increase.  In the face of public concerns, to say that there is {\it nothing} to worry about is both completely meaningless and impossible.  To say that the changes might call for increased caution is reasonable and supported by the data.  In this case, the secular trend of the frequency of larger shocks was a clear sign that there {\it might} be a change -- note, not an event -- in course.  

\section{Communication of risk}

If all this background seems obvious, it is not.   This was not communicated to the public.\footnote{A classic report by the American National Academy of Sciences is still as valuable as when it was written in 1989,   {\it Improving Risk Communication} (accessible online through http://www.nap.edu/catalog.php?record\_id=1189}

\subsection{The message}

Consider one press reaction to the l'Aquila Six verdict.  Citing the prediction by California seismologists that there was a 95\% chance of an earthquake occurring in a specific portion of the San Andreas fault, one of the most dangerous sites in North America for devastating  tremors, within eight years, one commentator noted ``and they were wrong''.   What does that remark mean?  Nothing!  They neither were, nor could be, wrong.  If something has a {\it chance} of happening it does not mean it {\it will}.  And 95\%, while it seems a good outcome for an election, is hardly reassuring from any statistical point of view.  It is, in fact, a statement of uncertainty and could just as easily be quantified as ``some'' chance.  But for quantification, it is what comes out of predictive models.  The comment, instead, made the {\it correct} scientific statement into something it was neither intended to be nor could be, a certainty.\footnote{For an excellent resource letter, now rather old (2006), see 
http://www.colorado.edu/hazards/publications/informer/infrmr2/pubhazbib.pdf}

If an event cannot be predicted with certainty, neither can its opposite be certain -- that something will {\it not} happen.  Probability assessments of risk run both ways.  Therefore it is imperative to properly nuance, with explanation in {\it non-technical language} the context for any statement.  Reporters often behave stupidly at this point, impatient and bored with details they don't want to understand.   It is far easier to ask ludicrously direct yes/no questions than wait for a reasoned reply.    Nonetheless, it is the scientists' obligation to persist in forcing the proper telling of their story, to insist that they be correctly quoted and to vigorously hound -- and correct -- those who distort their message.

\subsection{The method}

The scientific method, such as it is, is founded in doubt and anomaly.  Doubt in that any results must be scrutinized, not believed.  This also extends to the individual scientist.  If the ideas of others are to be questioned, so must our own.  And when we make a statement, whether about a research result or -- as in this case -- a potentially important public issue, we must do so with the conviction that we are possibly wrong.   No matter how expert a scientist may be, the world is neither linear nor simple and expertise in one area may be blind to knowledge and even contradictions in another.  That's where anomaly enters.   We never see the world `as it is'', that's why we have models.  These function as a way of ordering phenomena.  But they also function as filters, even blinkers.  The models may, if not used cautiously and skeptically, become the reality.  This is certainly true for the general public that knows only the pronouncements, the results, without knowing the reasoning behind it.  That's as much the fault of the scientists as it is the media.  It isn't merely a question of ''science education'' or even ''public outreach''.  The former can degenerate into ''literacy'' (whatever that has come to mean) or almost creed.  The latter all too often merely publicity, an attempt to propagandize and entertain the non-scientists with ``wonderful discoveries''.   Instead, critical thinking -- doubt -- is often missing from either sphere, even less present when the communication and teaching are left to interpreters who are themselves specialists only in ``communications''.  We have seen the disastrous effects of this in business, the ``creed'' of the CEO who knows how to manage without needing to know the why or how for the entity being managed.  To be aware that conditions are departing from the norm, or from expectations requires analysis and testing, debate and skepticism.

\subsection{Cases: poor or false}

Look at the reporting on the predictions in Italy during the storms of the week of 8 to 12 November 2012.   This was {\it not} the ``cousin of hurricane Sandy'', nor was {\it that} storm a {\it frankenstorm}.  The press, sensationalizing because it's their ``craft'', created an expectation far beyond the reality and was irresponsible.   This is not a {\it freedom of the press} issue.  Those who are the intermediaries between the sources (``experts'') and the receivers (``the public'' and officials responsible for decision-making) can only interpret potential risks based on the information provided and if that channel is distorted they cannot calibrate that out.  In communication theory, this comprises two effects: noise and distortion.  Noise is whatever masks the signal but, with a sufficient repetition (for instance) the original information is recoverable.  Distortion is far more insidious; it systematically changes the information content so the original signal cannot be reconstructed.  This second effect risks the production of a back-reaction when little actual damage is suffered or the event turns out to be less than ``predicted'' (the quotes are important).  The next prediction may be seen as unreliable and, consequently, ignorable.  

For the l'Aquila Six, the issue was the contrary.  The public was told by the spokesperson for the commission whose task was to evaluate the possibility of risk that there was {\it nothing} to worry about, that there was no indication of anything unusual.  More seriously, a false reason for calm was introduced into the communication, the {\it distortion} that the public could not evaluate.  But there was, in principle, a check on this: {\it the scientists on the commission, who were in a position to verify the information being transmitted to the public, could have spoken up immediately and provided the relevant and necessary calibration to an otherwise uniformed audience}.  They did not.   The prosecution and verdict were for that collective negligence.   Nobody else was in a position to challenge the statements made in the press conference, no one else had seen all of the details (in theory, at this stage-- mid-November 2012 -- we do not know the sum of evidence examined).   The minutes were written one week {\it after} the event and contained nothing specific, quantitative, or precise.  More damning was the preamble that stated that the {\it greatest experts} were convened to assess the risk.  This displays a tremendous {\it huberis} on the part of the communicators who drafted the document and the committee by giving their consent with their signatures.   It was designed to further the distortion of the signal.\footnote{see e.g. Showstack, R. 2012, {\it EOS}, {\bf 93}, 455, with links for the American Geophysical Union.}  

Testimony at the trial cited these calming affirmations of the commission as the basis for decisions that led to tragic results.   Residents of l'Aquila testified that they changed their habitual reactions to seismic activity based on the pronouncements of the commission.  While nothing might have happened, a quake might {\it not} have occurred, that is contrafactual.  It did.  The authorities might have been considered alarmists had there been an advisory of caution to no effect, but nothing would have resulted from that and in collective memory something would have been taken away from the event.  If both reporters and communicators at all levels had been careful to qualify their too firm pronouncements, to {\it explain} the context of the advisory, it is highly unlikely that there would have been a negative long-term effect in the credibility of such predictions.  But again, that is counterfactual in the specific case of the l'Aquila catastrophe.  Other events have been handled according to this model procedure -- volcanic activity, hurricane risk, flood warnings, tsunami warnings (think of Hawaii, where tsunami warnings are routinely received by the public with proper understanding and response) -- and they do not meet sketicism or cynicism on the part of the public.  

\section{Coda}

If something highly improbable, it is not absolutely impossible.\footnote{Adams, D. 1980, {\it The Hitchhiker's Guide to the Galaxy} (NY: Harmony Books)}  In the communication of risk, sufficient {\it  evidence and background} must be provided to permit the public, whoever they are, to understand the nuances of any prediction.  All predictions are provisional, and must be continually reviewed and revised in light of new information and this too must be openly communicated.  

\subsection{Appendix: some useful websites}

Here we include a small sample of the commentary now available on the web.  These reports have all appeared long after the verdict and reflect some of the``rethinking''.

\begin{itemize}
\item http://www.crikey.com.au/2012/10/26/why-scientists-should-sometimes-be-on-trial/

\item http://www.iaspei.org/news\_items/laquila\_IASPEI\_press\_release\_final.pdf

\item http://www.nature.com/news/l-aquila-verdict-row-grows-1.11683

\item http://blogs.scientificamerican.com/guest-blog/2012/10/22/the-laquila-verdict-a-judgment-not-against-science-but-against-a-failure-of-science-communication/

\item http://sciencepolicy.colorado.edu/admin/publication\_files/2002.22.pdf
\item http://www.washingtonpost.com/blogs/worldviews/wp/2012/10/24/the-deeper-issues-behind-italys-conviction-of-earthquake-scientists/

\item http://beforeitsnews.com/science-and-technology/2012/10/mischaracterizations-of-the-laquila-lawsuit-verdict-2484838.html

\item http://www.guardian.co.uk/science/controversiesinscience
\end{itemize}

\end{document}